\documentclass[aip,apl,twocolumn,10pt,showpacs,altaffillsymbol,superscriptaddress,longbibliography]{revtex4-1}
\pdfoutput=1
\usepackage{graphicx}
\usepackage{amsfonts}
\usepackage{amsmath}
\usepackage{natbib}

\begin{document}

\title{High-resolution error detection in the capture process of a single-electron pump}

\author{S.~P.~Giblin}
\author{P.~See}
\author{A.~Petrie}
\altaffiliation[Current address: ]{No affiliation}
\author{T.~J.~B.~M.~Janssen}
\affiliation{National Physical Laboratory, Hampton Road, Teddington, Middlesex TW11 0LW, United Kingdom}
\author{I.~Farrer}
\altaffiliation[Current address: ]{Department of Electronic and Electrical Engineering, University of Sheffield, Mappin Street, Sheffield, S13JD, United Kingdom}
\author{J.~P.~Griffiths}
\author{G.~A.~C.~Jones}
\author{D.~A.~Ritchie}
\affiliation {Cavendish Laboratory, University of Cambridge, J J Thomson Avenue, Cambridge CB3 0HE, United Kingdom}
\author{M.~Kataoka}
\affiliation{National Physical Laboratory, Hampton Road, Teddington, Middlesex TW11 0LW, United Kingdom}

\email[stephen.giblin@npl.co.uk]{Your e-mail address}

\date{\today}

\begin{abstract}
The dynamic capture of electrons in a semiconductor quantum dot (QD) by raising a potential barrier is a crucial stage in metrological quantized charge pumping. In this work, we use a quantum point contact (QPC) charge sensor to study errors in the electron capture process of a QD formed in a GaAs heterostructure. Using a two-step measurement protocol to compensate for $1/f$ noise in the QPC current, and repeating the protocol more than $10^{6}$ times, we are able to resolve errors with probabilities of order $10^{-6}$. For the studied sample, one-electron capture is affected by errors in $\sim30$ out of every million cycles, while two-electron capture was performed more than $10^6$ times with only one error. For errors in one-electron capture, we detect both failure to capture an electron, and capture of two electrons. Electron counting measurements are a valuable tool for investigating non-equilibrium charge capture dynamics, and necessary for validating the metrological accuracy of semiconductor electron pumps.
\end{abstract}

\pacs{1234}

\maketitle

Laterally-gated semiconductor quantum dots (QDs) have become a paradigmatic system for studying and manipulating charge carriers in tunable confining potentials. In one example, a dynamically-gated QD can pump electrons from a source to a drain electrode\cite{kouwenhoven1991quantized,blumenthal2007gigahertz,kaestner2008single,fujiwara2008nanoampere}, generating an accurate quantized current with metrological application as a new realisation of the ampere \cite{giblin2012towards,rossi2014accurate,bae2015precision,stein2015validation}. In this application, the accuracy of electron trapping and ejection in the non-adiabatic regime are of key importance. We wish to transfer $n$ electrons in response to a single gate voltage cycle, with an error rate less than $10^{-7}$. Recently, DC current measurements \cite{stein2015validation} have shown that $\sim 6\times 10^8$ electrons per second can be transferred through a QD with an \textit{average} error rate as low as $2 \times 10^{-7}$. To probe the \textit{individual} probabilities $P_{\text{n}}$, mesoscopic charge detectors need to be used to measure the number of electrons pumped onto charge sensing islands \cite{fricke2013counting,fricke2014self,yamahata2014accuracy,tanttu2015electron}, with the lowest reported error probability, $10^{-4}$, achieved in a silicon QD pump \cite{yamahata2014accuracy}. The error counting measurements on GaAs pumps \cite{fricke2013counting,fricke2014self} were performed in zero magnetic field using superconducting single electron transistor (SET) charge detectors, and were therefore not able to access the regime of high accuracy pumping attained in fields $>10$~T \cite{giblin2012towards,bae2015precision,stein2015validation}. Precise single-electron counting tests of semiconductor QD pumps in the high-field, high-accuracy pumping regime are clearly desirable to validate the DC pumping measurements, and also to approach the benchmark error probability of $\sim 10^{-8}$ demonstrated for a slow adiabatic metallic pump \cite{keller1996accuracy}. 

In this work we focus on measuring $P_{\text{N}}$, the probability of loading the QD with $N$ electrons (distinguished from $P_{\text{n}}$, the probability of pumping $n$ electrons). There is good experimental evidence that the ejection of electrons to the drain can be accomplished with a sufficiently low error rate to satisfy metrological criteria \cite{miyamoto2008escape}, and furthermore theoretical treatments of the QD pumping process \cite{fujiwara2008nanoampere,kashcheyevs2010universal,kashcheyevs2012quantum,kashcheyevs2014modeling} have shown that $n$ is determined during the loading stage. We use a quantum point contact (QPC) charge detector strongly coupled to the QD, to probe the number of loaded electrons. The use of a QPC instead of SET detector \cite{fricke2013counting,fricke2014self} allows operation in strong magnetic fields. Electrons can be loaded with fast voltage waveforms, as used in high-speed pumping experiments \cite{giblin2012towards,stein2015validation} and then probed on millisecond time-scales required for high-fidelity readout. We use a 2-stage measurement protocol to suppress the effect of $1/f$ noise in the QPC due to non-equilibrium charged defects, and we achieve a sufficiently high charge detection fidelity (probability of the charge detection yielding the right answer for $N$), to probe loading probabilities at the $10^{-6}$ level. Finally, we demonstrate good qualitative agreement between electron-detection measurements of $N$, and $\langle n \rangle$ extracted from the pumped current measured in a separate experiment.

\begin{figure}[!]
\includegraphics[width=8.5cm]{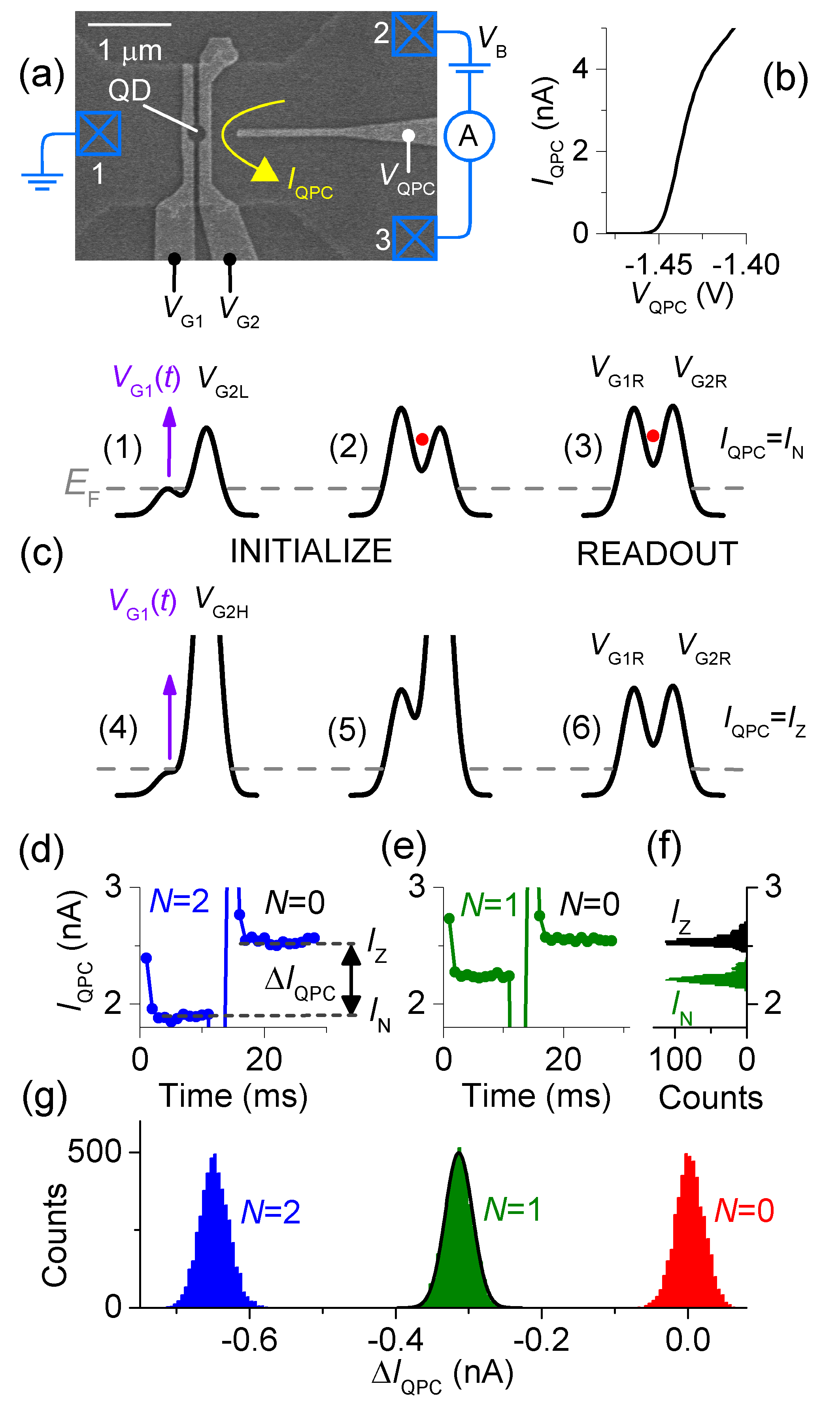}
\caption{\textsf{(a): SEM image of device and electrical connections to the measurement circuit. The crossed boxes numbered $1-3$ indicate Ohmic contacts to the 2-DEG. The ammeter could also be connected to contact 1 to measure the pumped current (see text). (b): QPC pinch-off characteristic with $V_{\text{B}}=1$~mV. (c) Frames (1-3): schematic potential landscape showing initialization of the QD with one electron. Frames (4-6): initialization of the QD with zero electrons. Frames (1-6) illustrate one measurement cycle. (d): example of raw QPC current data for one measurement cycle, for the case of two electrons loaded into the QD, illustrating the difference signal $\Delta I_{\text{QPC}}$. (e): as (d), but one electron is loaded into the QD. (f): Histograms of the QPC currents corresponding to zero and one electron, obtained from 950 measurement cycles spaced evenly over $15$~hours. Plots (d-f) share the same y-axis. (g) Histograms of $\Delta I_{\text{QPC}}$ for $3$ sets of $5000$ cycles, with $V_{\text{G2L}}$ set to $-0.65$~V, $-0.616$~V and $-0.59$~V, to load respectively $0$, $1$ and $2$ electrons into the QD. Solid black line: Gaussian fit to the $N=1$ data. $B=12$~T for all data in this figure.}}
\label{fig:fig1}
\end{figure}

A scanning electron microscope image of a device similar to the one used in this study is shown in Fig. 1a. The device is fabricated on a GaAs/Al$_{x}$Ga$_{1-x}$As wafer using optical and electron-beam lithography and wet-chemical etching \cite{blumenthal2007gigahertz}. A 2-dimensional electron gas (2-DEG) with density $2.3 \times 10^{11}$~cm$^{-2}$ and mobility $2.28 \times 10^{6}$~cm$^{2}$ V$^{-1}$ s$^{-1}$is formed $90$~nm below the surface. The QD pump is similar to those used in previous high-precision measurements of the pumped current \cite{giblin2012towards}: voltages $V_{\text{G1}}$ and $V_{\text{G2}}$ applied to the entrance (left) and exit (right) gates form a QD in the cut-out region between the gates. In this work, we introduce a third gate, biased with voltage  $V_{\text{QPC}}$, to form a QPC close to the QD. When the channel between the QPC gate and the exit gate is biased close to pinch-off (Fig. 1b), the conductance of the channel probes the local charge environment and therefore the number of electrons in the QD \cite{field1993measurements}. We applied a source-drain bias voltage $V_{\text{B}}=1$~mV across the QPC channel and measured the current $I_{\text{QPC}}$ using a room temperature transimpedance amplifier with gain $10^6$ V/A and $400$~Hz bandwidth. The amplifier output was continuously digitized by an integrating voltmeter with 1 ms aperture. All measurements were performed in a sorbtion-pumped helium-3 cryostat, at base temperature $\sim 300$~mK. A magnetic field was applied perpendicular to the plane of the 2-DEG to enhance the current quantization \cite{giblin2012towards,fletcher2012stabilization,bae2015precision,stein2015validation}.

Our measurement protocol is illustrated by schematic potential diagrams in Fig. 1c. The QD is initialized (frame 1) by applying a voltage $V_{\text{G1}}(t)$ to the entrance gate, raising the entrance barrier above the Fermi level of the source electrode and trapping $N$ electrons in the QD. $N$ is tuned by the fixed exit gate voltage $V_{\text{G2L}}$. Two types of entrance gate ramp are used in the experiments, denoted 'slow' (Fig. 2a) and 'fast' (Fig. 2d), differing in rise-time by a factor $\sim 10^5$. The entrance gate ramp $V_{\text{G1}}(t)$ terminates at $V_{\text{G1}}=V_{\text{G1R}}$ (frame 2), with the electrons trapped in a deep potential well. Finally, $V_{\text{G2}}$ is adjusted to its readout value $V_{\text{G2R}}$ (frame 3), and the QPC current $I_{\text{N}}$ is measured following a $4$~ms delay to reject transient effects. This final adjustment is a convenience which allows readout of the charge state at fixed pump gate voltages $[V_{\text{G1R}},V_{\text{G2R}}]$ independently of the tuning parameter $V_{\text{G2L}}$ \cite{mcneil2011demand}.

To ensure high detection fidelity in the presence of $1/f$ noise, we reference the QPC current ($I_{\text{N}} \propto -N$ for small $N$) to a second current $I_{\text{Z}}$.  $I_{\text{Z}}$ is measured with the QD initialised to a known $N=0$ state, by raising the entrance barrier with the exit gate set to a large negative value $V_{\text{G2H}}$ (frame 4). $N$ is then determined from the difference $\Delta I_{\text{QPC}}=I_{\text{N}}-I_{\text{Z}}$. The series of frames (1-6) illustrate one measurement cycle. Raw QPC current data is shown in Figs.~1d and 1e for cases $N=2$ and $N=1$ respectively. Fig. 1f shows histograms of $I_{\text{N}}$ (green, lower peak) and $I_{\text{Z}}$ (black, upper peak) for a set of measurement cycles loading one electron, spread over $15$ hours. The effect of $1/f$ noise is visible as asymmetric broadening of the peaks, and the fidelity of measuring $N$ using the $I_{\text{N}}$ data alone is estimated as $\sim0.999$. Fig.~1g shows histograms of $\Delta I_{\text{QPC}}$ obtained from $3$ runs of $5000$ cycles each, with $V_{\text{G2L}}$ set to load approximately $0$, $1$ and $2$ electrons using the slow loading waveform. The histograms form $3$ widely-separated peaks, which we identify with $N=0,1,2$, which fitted well to normal distributions (shown for $N=1$) with standard deviation $\sigma =20$~pA. For these relatively short calibration runs of $5000$ cycles, there were no out-liers inconsistent with the normal distribution. Applying a simple thresholding algorithm for determining $N$ from $\Delta I_{\text{QPC}}$, the intrinsic probability $P_{\text{fail}}$ of measuring the wrong value of $N$ is given by $1-erf(\Delta I_{\text{1e}}/ 2\sqrt{2} \sigma)$, where $\Delta I_{\text{1e}}=320$~pA is the separation between peaks. $P_{\text{fail}}$ is weakly dependent on the magnetic field $B$, and for the values of $B$ used in this study, $P_{\text{fail}}<10^{-8}$ which is much smaller than the expected statistical uncertainty in $P_{\text{N}}$.

\begin{figure}[!]
\includegraphics[width=8.5cm]{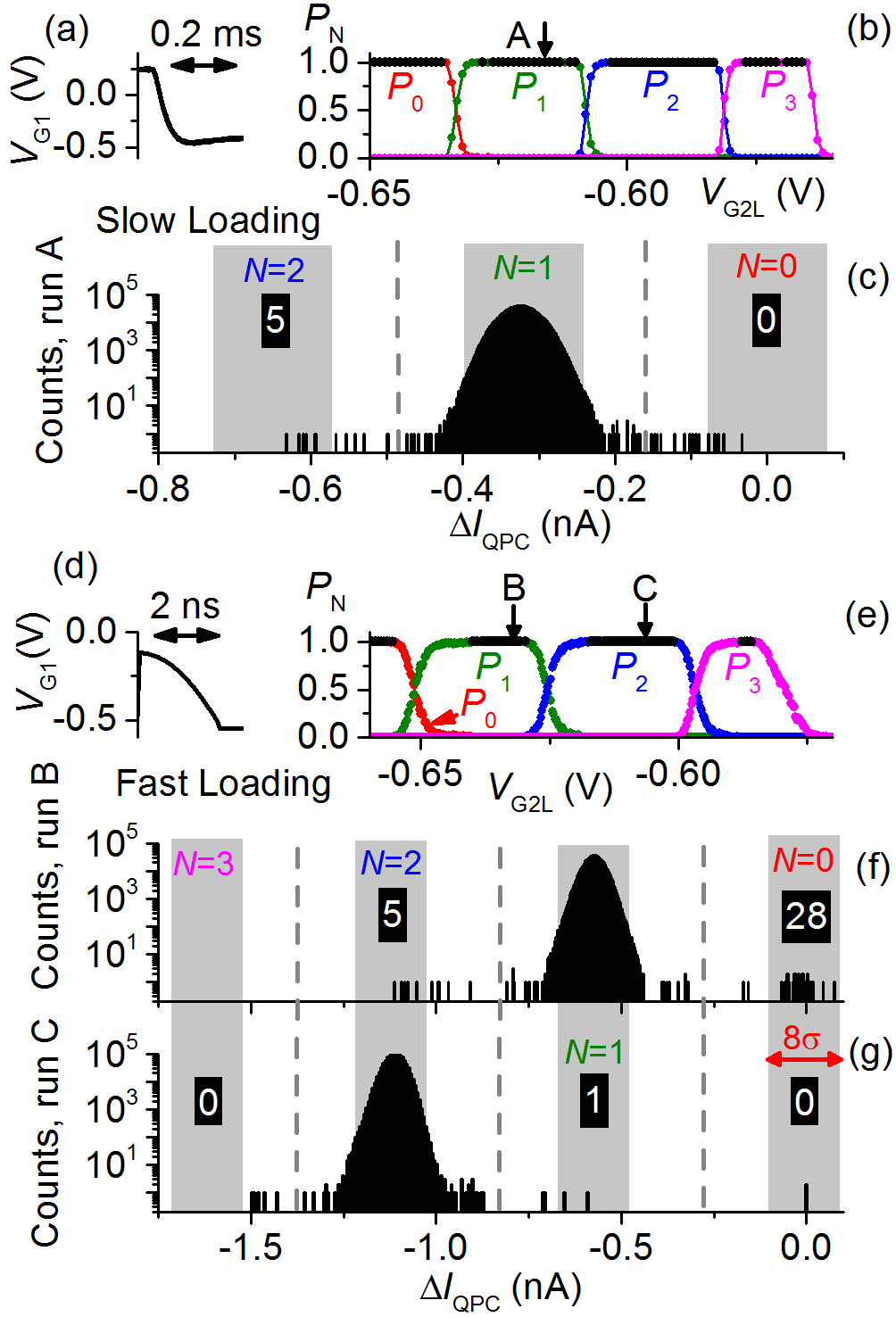}
\caption{\textsf{(a): Time dependence of the slow entrance gate loading waveform, measured on a sampling oscilloscope. (b) Loading probabilities for up to $3$ electrons as a function of exit gate voltage using the slow load waveform. $B=12$~T. Black data points denote values of $V_{\text{G2L}}$ where all the cycles yielded the same $N$. (c) Log-scale histogram of $\Delta I_{\text{QPC}}$ for $1847154$ cycles using the slow waveform, at the exit gate voltage denoted by the vertical arrow marked 'A' in graph (b). Vertical dotted lines show the thresholds for discriminating $0$, $1$ and $2$ electrons, and shaded grey boxes show the $\pm4\sigma$ width of distributions obtained from short calibrations runs similar to Fig. 1(g). (d): as (a), but showing the fast loading waveform. (e): as (b), but using the fast waveform at $B=10$~T. (f): as (c), but using the fast waveform, $1045310$ cycles at the gate voltage indicated by 'B' in (e). (g): $1149457$ cycles at the gate voltage indicated by 'C' in (e). Plots (f) and (g) share the same x-axis, and $B=10$~T. Boxed numbers in plots c,f,g indicate the number of counts for the relevent $N$ (see text).}}
\label{fig:fig2}
\end{figure}

In Fig. 2b and 2e we show $P_{\text{N}} (V_{\text{G2L}})$ for $0 \leq N \leq 3$, obtained from sets of $500$ cycles for each $V_{\text{G2L}}$ using the slow and fast loading waveforms respectively. As expected from DC current measurements with the pump in a magnetic field \cite{giblin2012towards,fletcher2012stabilization,bae2015precision,stein2015validation}, there are a series of wide plateaus where one value of $N$ dominates the loading statistics. This is highlighted by color-coding data points black when all the cycles yielded the same value of $N$. Comparable data has been presented previously \cite{fricke2013counting}, but at zero magnetic field where the regime of very accurate loading could not be accessed. We calculated $\langle N\rangle = \sum NP_{\text{N}}$ for the fast loading data, and fitted it to both the decay cascade model \cite{fujiwara2008nanoampere,kashcheyevs2010universal} (Fig.4c, inset) and also a thermal equilibrium model \cite{fricke2013counting,yamahata2014accuracy}. The decay cascade model yielded a better fit, as was also found at the much lower temperature of $25$~mK \cite{fricke2013counting}. The plateaus are considerably sharper using the slow loading pulse, consistent with pumped current measurements in which the rise time (proportional to the inverse of the pumping frequency) was varied in the $1-10$~ns range\cite{giblin2012towards}. Our technique allows the investigation of rise times over many orders of magnitude, including those too slow to generate a measurable pumped current. We note that the decay cascade model does not predict any rise-time dependence to the plateau shape \cite{fujiwara2008nanoampere,kashcheyevs2010universal}, and the deterioration of the plateaus as the rise time is reduced \cite{giblin2012towards, stein2015validation} does not currently have a satisfactory explanation. 

To assess the loading accuracy on the plateaus, we performed a number of long runs, comprising $\sim 10^6$ cycles at fixed $V_{\text{G2L}}$ chosen to preferentially load always the same desired $N$, $N_{\text{DES}}=1,1,2,1$ for runs A, B, C and D respectively. Only run A used the slow pulse. The quantity of interest is the number of times $N\neq N_{\text{DES}}$, an event we refer to as an 'error'. Histograms of $\Delta I_{\text{QPC}}$ for three of these runs, denoted A, B and C, are presented in Figs. 2c, 2f and 2g, plotted on a log scale to highlight rare events. For each run, most of the counts comprise a Gaussian peak corresponding to $N=N_{\text{DES}}$, but for run B there is also a small $N=0$ peak. For all the runs, there is a background of events which are not statistically compatible with any $N$ state. Examination of the raw $I_{\text{QPC}}(t)$ data for these events showed that most of them could be attributed to TLF events well known to occur in our type of GaAs device \cite{cobden1991noise}. $P_{\text{N}}, \{ N \neq N_{\text{DES}} \}$ was evaluated from the small number of counts, indicated on Figs. 2c,f,g, with $\Delta I_{\text{QPC}}$ within $\pm 4\sigma$ of the expected value for a given $N \neq N_{\text{DES}}$, which could not be attributed to TLF events. From this data, we calculated the most probable values for $P_{\text{N}}$ and the asymmetric statistical uncertainties (half-width at half maximum) using the binomial expression. For run B, $P_{\text{0}}=(26.7^{+6.4}_{-5.5}) \times 10^{-6}$, $P_{\text{2}}=(4.8^{+3.0}_{-2.1}) \times 10^{-6}$ and for run C, $P_{\text{0}}=P_{\text{3}}=(0^{+0.7}_{-0}) \times 10^{-6}$, $P_{\text{1}}=(0.9^{+1.3}_{-0.8}) \times 10^{-6}$. $P_{\text{N}}$ for runs A-D are plotted as open points in Fig. 3a and 3b (the error bars are smaller than the data points), along with the data of Figs 2b and 2e re-plotted on a log scale for comparison (solid points). For $N_{\text{DES}}=1$, the measured error rates are almost an order of magnitude larger than predicted by fits to the decay-cascade model (solid line in Fig. 3b)\cite{fujiwara2008nanoampere,kashcheyevs2010universal}. Additionally, the co-existence of $N=0$ and $N=2$ errors at the same pump operating point highlights the importance of electron counting measurements, as these errors would partially cancel in an average current measurement. This data also shows that caution should be exercised in using theoretical fits to low-resolution data \cite{giblin2012towards,rossi2014accurate,bae2015precision,stein2015validation} as a method of predicting the accuracy of an tunable-barrier electron pump on a quantized plateau.

\begin{figure}[!]
\includegraphics[width=8.5cm]{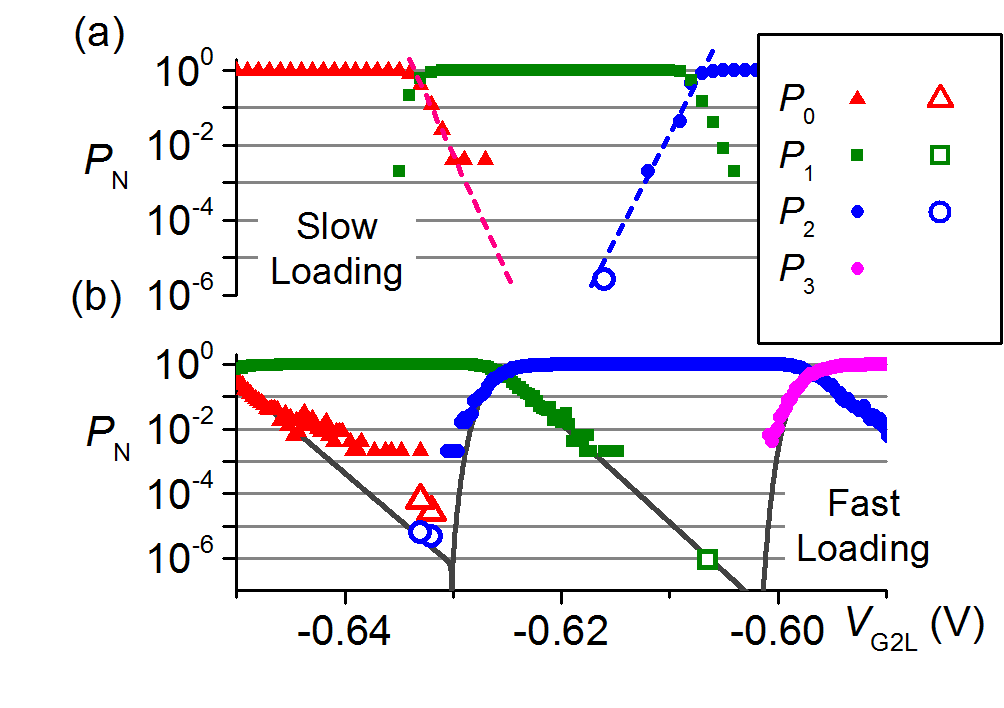}
\caption{\textsf{(a): solid points: the $P{_\text{N}}$ data of Fig. 2(b) re-plotted on a log scale. Open point: $P_{\text{2}}$ calculated from the data of run A. Dotted lines are guides to the eye. (b): solid points: data of Fig. 2(e) re-plotted on a log scale. Open points: $P_{\text{0}}$, $P_{\text{1}}$ and $P_{\text{2}}$ calculated from the data of runs B-D. Error bars (see text) are smaller than the plotted points. Solid lines are theoretical fits to the $\langle N \rangle=1$ and $\langle N \rangle=2$ plateaus (See text).}}
\label{fig:fig3}
\end{figure}

Finally, we compare $\langle N\rangle$ computed from the data set of Fig 2e, with the normalised pump current $I_{\text{P}}/ef$ when the load waveform was immediately followed by a pulse to eject the trapped electrons to the drain (pump waveform Fig. 4b) \cite{giblin2012towards,stein2015validation}. The fast load waveform is shown in Fig. 4a for comparison. The repetition frequency of the pump waveform was $280$~MHz, generating a current $I_{\text{P}}\sim 45$~pA measured by connecting the ammeter to contact $1$ in Fig. 1a, and grounding contacts $2$ and $3$. The loading parts of the two waveforms had the same $V_{\text{G1}}(t)$ profile (grey boxes in Fig. 4a,b). In Fig. 4c, $\langle N\rangle$ and $I_{\text{P}}/ef$ are plotted as a function of exit gate voltage (for the case of the pumping data, the x-axis is the constant DC voltage applied to the exit gate). The reasonable agreement between the two measurements suggests that the electron loading experiment is probing the same dynamical process which determines the DC current in pumping experiments, and furthermore suggests that the wide quantised pumping plateaus seen here, and in previous studies of QD pumps \cite{kaestner2008single,fujiwara2008nanoampere,giblin2012towards,rossi2014accurate,bae2015precision,stein2015validation}are indeed due to transport of the same number of electrons in each pumping cycle. On the other hand, close examination of Fig. 4c shows that the transitions between plateaus are slightly broader in the loading experiment. This could be evidence for back-action of the QPC on the electron loading process, since the QPC source-drain bias voltage was not present during the pumping experiment. The irregularity in the transition from $\langle N \rangle=0$ to $\langle N \rangle=1$ clearly visible in Fig. 3b was not present in the DC current measurements. Further studies will clarify the back-action of the QPC on the dynamic QD.

\begin{figure}[!]
\includegraphics[width=8.5cm]{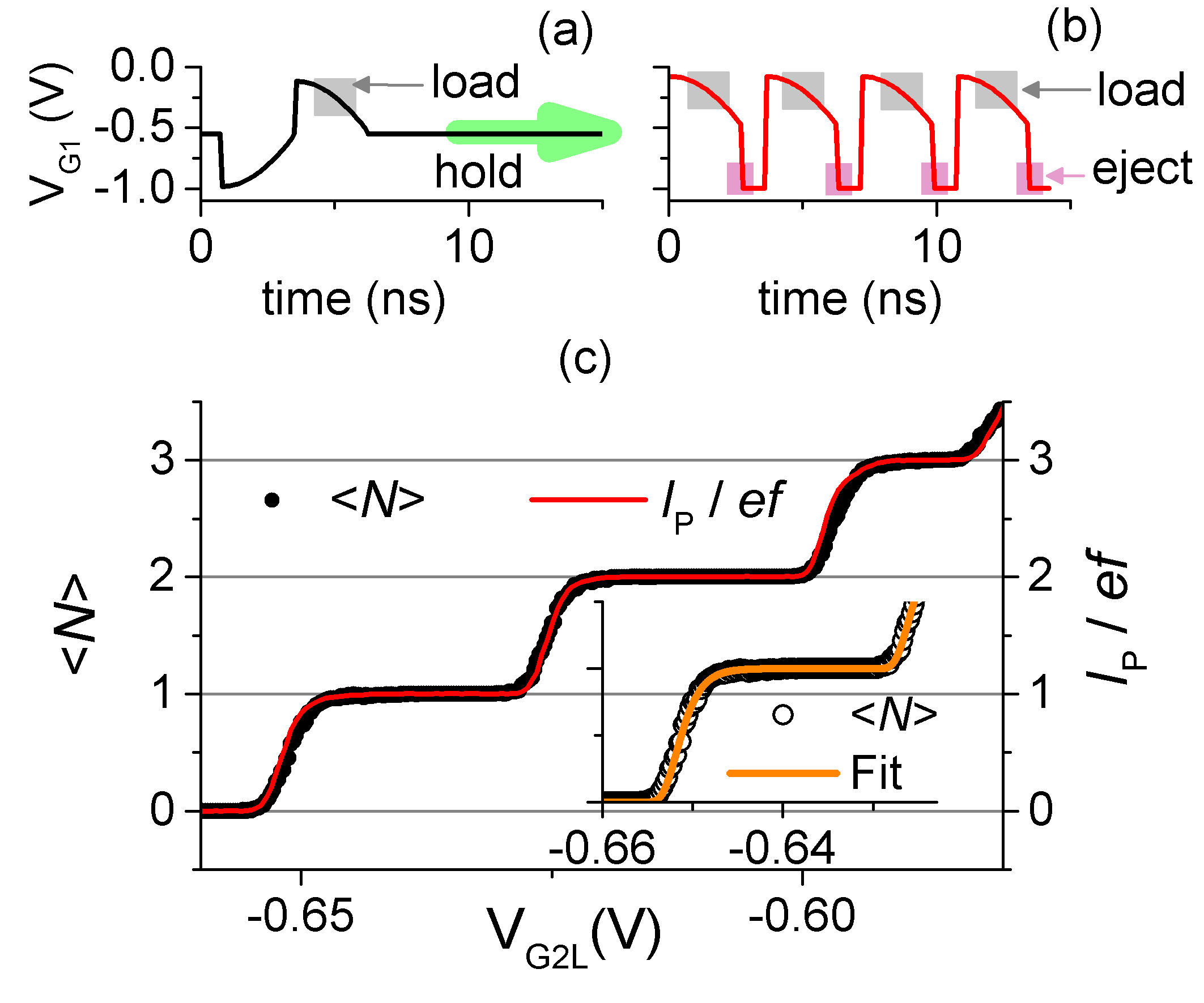}
\caption{\textsf{(a): Time dependence of the fast waveform used to load electrons. The negative-going part at the beginning of the waveform ejects any trapped electrons to the drain, but plays no role in the loading experiment. (b): Pumping waveform. (c): comparison of average number of electrons loaded (points) with pumped current (line) in loading and pumping experiments respectively. The inset shows the portion of the loading data from the main figure with a fit to the decay-cascade model.}}
\label{fig:fig3}
\end{figure}

In summary, we have presented a simple device architecture and measurement protocol for studying directly the loading statistics of a dynamic QD with high precision. By incorporating a reference measurement using a known charge state, $1/f$ noise in the QPC detector is compensated, and  the number of electrons in the QD can be measured with an intrinsic fidelity exceeding $1-10^{-8}$. This is a significant step towards validating semiconductor QD pumps as metrological current sources. Our device can also be used to study the QD initialization process over many orders of magnitude in barrier rise time, without constraints imposed by the need to measure a small DC current. This will help to clarify the role of quantum non-adiabaticity \cite{kataoka2011tunable,kashcheyevs2012quantum,kashcheyevs2014modeling} in the formation of the QD. Improvements to the control and readout electronics will allow $\sim 10^7$ cycles in a reasonable $12$~hour experimental run, reducing the statistical uncertainty in the measurement of small probabilities, and experimental precautions against TLFs, for example biased cool-down \cite{PioroLadri2005origin}, should remove these unwanted artifacts in future experiments.

\begin{acknowledgments}
We would like to thank Akira Fujiwara, Joanna Waldie and James Frake for useful discussions and Stephane Chretien for assistance with statistical analysis. This research was supported by the UK department for Business, Innovation and Skills and within the Joint Research Project 'Quantum Ampere' (JRP SIB07) within the European Metrology Research Programme (EMRP). The EMRP is jointly funded by the EMRP participating countries within EURAMET and the European Union.
\end{acknowledgments}

\bibliography{SPGrefsEDet}
\bibliographystyle{apsrev}

\end{document}